\documentstyle[emulateapj]{article} 
 
\slugcomment{({\it ApJ Letters} accepted)}
 
\lefthead{J.Koda, Y.Sofue and K.Wada}
\righthead{Unified Scaling Law}

\newcommand{\Msun}{M_{\odot}}
\newcommand{\kpc}{{\rm \, kpc}}
\newcommand{\mpc}{{\rm \, Mpc}}
\newcommand{\magni}{{\rm \, mag}}
\newcommand{\kmps}{{\rm \, km \, s^{-1}}}
\newcommand{\magpsec}{{\rm \, mag \, arcsec^{-2}}}
\newcommand{\kmpspmpc}{{\rm \, km \, s^{-1} \, Mpc^{-1}}}

\begin{document}
\title{A Unified Scaling Law in Spiral Galaxies}

\author{Jin Koda$^1$, Yoshiaki Sofue$^1$ and Keiichi Wada$^2$}
\affil{1.Institute of Astronomy, University of Tokyo, Mitaka, Tokyo 181-8588, Japan\\
2.National Astronomical Observatory, Mitaka, Tokyo 181-8588, Japan
}

\begin{abstract}
We investigate the origin of a unified scaling relation in spiral galaxies.
Observed spiral galaxies are spread on a plane in the three-dimensional
logarithmic space of luminosity $L$, radius $R$ and rotation velocity $V$.
The plane is expressed as $L \propto (V R)^{\alpha}$ in $I$-passband,
where $\alpha$ is a constant. On the plane, observed galaxies are distributed
in an elongated region which looks like the shape of a surfboard.
The well-known scaling relations, $L-V$ (Tully-Fisher relation), $V-R$
(also the Tully-Fisher relation) and $R-L$ (Freeman's law), can be
understood as oblique projections of the surfboard-like plane into 2-D spaces.
This unified interpretation of the known scaling relations should be a clue
to understand the physical origin of all the relations consistently. 
Furthermore, this interpretation can also explain why previous studies could
not find any correlation between TF residuals and radius.

In order to clarify the origin of this plane, we simulate formation
and evolution of spiral galaxies with the $N$-body/SPH method, including
cooling, star formation and stellar feedback. Initial conditions are set
to isolated 14 spheres with two free parameters, such as mass and angular
momentum. The CDM ($h=0.5$, $\Omega_0=1$) cosmology is considered as a test
case. The simulations provide the following two conclusions:
(a) The slope of the plane is well reproduced but the zero-point is not.
This zero-point discrepancy could be solved in a low density ($\Omega_0<1$)
and high expansion ($h>0.5$) cosmology.
(b) The surfboard-shaped plane can be explained by the control of galactic
mass and angular momentum.

\end{abstract}


\section{INTRODUCTION}
Luminosity $L$, radius $R$ and rotation velocity $V$ are basic parameters
for spiral galaxies. We have known the correlations between each two of
them: the log$L$-log$V$ (Tully \& Fisher 1977; TF), log$V$-log$R$
(also Tully \& Fisher 1977) and log$R$-log$L$ (Freeman 1970) correlations.
These scaling relations provide an observational benefit to measure galaxy
distances (e.g., Strauss \& Willick 1995; Giovanelli et al. 1997),
and also provide theoretical benchmarks to understand the structure,
formation and evolution of spiral galaxies (e.g., Dalcanton, Spergel \&
Summers 1997; Silk 1997; Mo, Mao \& White 1998).

There have been many efforts to search tighter correlations than these
three. In order to improve the accuracy of distance estimation,
a third-parameter effect on the TF relation, i.e. a correlation between
TF residuals and a third parameter, have been sought by many authors
(e.g., Willick et al. 1997; Courteau \& Rix 1999).
Most of them have concluded that the third-parameter effect may not be
crucial, while Willick (1999) has found a slight dependence of TF residuals
on surface brightness.
On the other hand, principal component analyses have suggested that
two parameters are necessary and sufficient to describe spiral galaxies
(see Djorgovski 1992, for a review), in contrast to stars which are described
by one parameter (mass). Kodaira (1989) has found that the correlation
among all the three parameters, log$L$, log$R$ and log$V$, is much tighter
than the correlations between each two of them.
Koda \& Sofue (2000) have recently found that spiral galaxies are distributed
on a surfboard-shaped plane in the 3-D space (log$L$, log$R$, log$V$).
The 2-D scaling relations ($L$-$V$, $V$-$R$, $R$-$L$) can be understood
uniformly as oblique projections of this surfboard-shaped plane.
Koda \& Sofue (2000) also argued that this unified scaling relation would
be produced through galaxy formation which is affected by galactic mass and
angular momentum.

Theoretically the importance of mass and angular momentum in the structure
of spiral galaxies has, of course, been discussed by many authors
(e.g., Fall \& Efstathiou 1980; Kashlinsky 1982).
Recently, the 2-D scaling relations ($L$-$V$, $V$-$R$, $R$-$L$) have been
discussed as the products of galaxy formation which is controlled by mass and
angular momentum (Dalcanton, Spergel \& Summers 1997; Mo, Mao \& White 1998;
Koda, Sofue \& Wada 2000). In this {\it Letter}, we discuss whether
the unified scaling relation (plane) in the 3-D space can also be
a product of mass and angular momentum.
We take the $N$-body/SPH approach which includes cooling, star formation
and stellar feedback (see Tissera, Lambas \& Abadi 1997;
Weil, Eke \& Efstathiou 1998; Steinmetz \& Navarro 1999; Elizondo et al. 1999;
Koda, Sofue \& Wada 2000), and consider the formation of 14 galaxies
with different masses and angular momenta.
The simulated galaxies show internal structures as observed in spiral
galaxies, e.g., the exponential density profile, flat rotation curve,
and distributions of stellar age and metallicity. Using these simulated
``spiral galaxies'', we try to confirm the origin of the unified scaling
relation.

\section{OBSERVATIONAL FACT} \label{sec:obs}
We briefly introduce the unified scaling relation in spiral galaxies.
Throughout this {\it Letter}, we use the data set presented by Han (1992),
which consists of member galaxies in 16 clusters.
All the sample galaxies in each cluster are assumed to be at the same distance
indicated by the systemic recession velocities of the host cluster,
which are measured in the CMB reference frame (Willick et al. 1995).
We assume $h=0.5$, where $h$ is the present Hubble constant in units of
$100 \kmpspmpc$.
In order to select exact members of a cluster, we reject galaxies whose
recession velocities deviate more than $1,000 \kmps$ from the mean velocity
of the cluster. We use total $I$-band magnitude $M_{\rm I} (\magni) $,
HI velocity width $W_{\rm 20} (\kmps)$ and face-on $I$-band isophotal radius
$R_{\rm 23.5} (\kpc)$. Final samples consist of 177 spiral galaxies.

When we consider the 3-D space of luminosity $L$, radius $R$ and rotation
velocity $V$, observed spiral galaxies are (i) distributed on a plane
as $L \propto (V R)^{1.3}$ and (ii) distributed in a surfboard-shaped region
on the plane (Koda \& Sofue 2000). Figure \ref{fig:schem} schematically
illustrates the situation with parameters of radius $\log R$, velocity $\log W$
and absolute magnitude $M_I$. Since the well-known 2-D scaling relations
($L-V$, $V-R$, $R-L$) can be understood uniformly as oblique projections
of this surfboard-shaped plane (Figure \ref{fig:schem}), we hereafter call
the plane {\it the scaling plane}. The upper panels of
Figure \ref{fig:comp} show the Tully-Fisher projection (left) and
the edge-on projection (right) of the scaling plane. The edge-on projection
has tighter correlation than the Tully-Fisher projection.
The same plane can be found in the data sets of Mathewson et al. (1992) and Courteau (1999) as well.
Note the $L-V$, $V-R$ and $R-L$ relations themselves may also be found as
the projections of a prolate (not thin plane) distribution 
in a 3-D space. However, the plane distribution unifies the scatters
of these three 2-D correlations as well.

In the 3-D space, observed galaxies are spread in the range of the order of
two for $L$, and the several factors for $R$ and $V$. Hence the scaling plane
has exactly the elongated (surfboard) shape. The primary and secondary axes
are schematically illustrated in Figure \ref{fig:schem}. We hypothesize
(a) that the 2-D distribution implies the existence of two dominant physical
factors in spiral galaxy formation, and (b) that one of them is more
dominant than the other because of the surfboard shape.

\section{NUMERICAL EXPERIMENT}

\subsection{Numerical Methods}
We simulate formation and evolution of spiral galaxies by the $N$-body/SPH
method similar to Katz (1992) and Steinmetz \& M\"uller (1994,1995).
We use a GRAPE-SPH code (Steinmetz 1996), a hybrid scheme of the smoothed
particle hydrodynamics and the $N$-body integration hardware GRAPE-3
(Sugimoto et al. 1990). This code can treat the gravitational and
hydrodynamical forces, radiative cooling, star formation and stellar feedback
(see Koda, Sofue \& Wada 2000 in details).

We take a phenomenological model of star formation. If a region is locally
contracting and Jeans-unstable, stars are formed at a rate $\dot{\rho_\star}
= c_\star \rho_{\rm gas}/ \max (\tau_{\rm dyn}, \tau_{\rm cool})$, where
$\rho_\star$, $\rho_{\rm gas}$, $\tau_{\rm dyn}$ and $\tau_{\rm cool}$ are
the local densities of stars and the gas, dynamical and cooling timescales,
respectively. We set $c_{\star}=0.05$.
We assume that massive stars with mass $m \geq 8 \Msun$ explode as type II
supernovae and release energy ($10^{51} {\rm erg}$), mass $(m-1.4\Msun)$
and metals ($16 \%$ of total released mass on an average;
see Nomoto et al. 1997a) into the surrounding
gas at a constant rate in the first $4 \times 10^{7} {\rm yr}$ from their
birth. They leave white dwarfs with mass $1.4 \Msun$ after the explosion.
And 15 \% of the white dwarfs are assumed to result in type Ia supernovae
(Tujimoto et al. 1995), which release energy ($10^{51} {\rm erg}$),
mass ($1.4 \Msun$) and metals ($100 \%$ of total released mass;
see Nomoto et al. 1997b) into the surrounding gas.
The number of the massive stars is counted with the initial mass function
(IMF) of Salpeter (1955) and we set the lower $m_l$ and upper mass $m_u$ of
stars to $(m_l, m_u) = (0.1 \Msun, 60 \Msun)$. The energy is released
into the surrounding gas as thermal energy.

\subsection{Initial Conditions}
We consider 14 homogeneous spheres which are rigidly rotating, isolated,
and overdense above the background field by $\delta \rho/\rho=0.25$.
The spheres follow the reduced Hubble expansion at $z=25$ in the CDM
cosmology ($\Omega_0=1$, $h=0.5$ and the rms fluctuation in $8 h^{-1} \mpc$
spheres $\sigma_8=0.63$). Small scale CDM fluctuations are superimposed
on the considered spheres.
We use the same realization (random numbers) of the fluctuations for
all the 14 galaxies. Two free parameters, total mass $M$ and spin parameter
$\lambda$, are $M = 8 \times 10^{11} \Msun$ ($\lambda=0.10, 0.08, 0.06$),
$4 \times 10^{11} \Msun$ ($0.08, 0.06, 0.04$),
$2 \times 10^{11} \Msun$ ($0.10, 0.08, 0.06, 0.04$), and
$1 \times 10^{11} \Msun$ ($0.10, 0.08, 0.06, 0.04$).
Since we consider collapses of isolated spheres, there is no infall of
clumps at low redshift which causes an extreme transfer of angular momentum
from baryons to dark matter (Navarro, Frenk \& White 1995;
Steinmetz \& Navarro 1999).

The gas and dark matter are represented by the same number of particles,
and their mass ratio is set to $1/9$ (Steinmetz \& M\"uller 1995). The mass
of a gas particle varies between $2.4 \times 10^{6}$ and
$1.9 \times 10^{7} \Msun$ according to the system mass considered.
The mass of a dark matter particle is between
$2.1 \times 10^{7}$ and $1.7 \times 10^8 \Msun$. Low resolution may cause
artificial heating due to two-body relaxation, however this range of
particle mass is small enough to exclude the artificial heating effect
(Steinmetz \& White 1997). The gravitational softenings are taken to
be $1.5 \kpc$ for gas and star particles, and $3 \kpc$ for dark matter.

\section{RESULTS}
We compute absolute magnitude $M_I$ of each ``spiral galaxy'' at $z=0$
with the simple stellar population synthesis models of Kodama \& Arimoto
(1997), and take the isophotal radius $R_{23.5}$ $ (\kpc)$ at the level $23.5
\magpsec$ in $I$-band. The line-width $W_{20}$ $(\kmps)$ is derived in
a way similar to observation by constructing a line-profile of gas,
and measuring the width at $20 \%$ level of a peak flux.
All the simulated galaxies have the exponential-light
profile and the flat rotation curve (see Koda, Sofue \& Wada 2000).

\subsection{The Scaling Plane of Simulated Galaxies}
In Figure \ref{fig:comp}, we compare the observed (upper panels)
and simulated (lower panels) distributions of spiral galaxies
in the TF projection (right panels) and edge-on (left panels)
projection of the scaling plane.
In the lower panels, the dotted lines represent the observed correlations
(as do the solid lines in the upper panels), and we shift the zero-point
of the solid lines to fit the simulations.
The ranges of the figures are shifted between the upper and lower panels
because of the systemic shift of simulated galaxies.
The lengths of the axes, however, are exactly the same and 
we can compare the slope and scatter between the upper and lower panels.

In this figure, we find the following three points:
(i) The slope and scatter of both correlations are well reproduced
in the simulation.
Note that the slope and scatter of $L-R$ and $R-V$ are also
consistent with the observations. (ii) The edge-on projection
of the simulated scaling plane shows a much better correlation than
the simulated TF projection, similar to the observations. The simulations
reproduce the slope and scatter of the scaling plane well.
(iii) However, the distribution of simulated galaxies is systematically
shifted from that of observed galaxies.

The systemic shift of the simulated distribution from the observed one
amounts to $(\Delta M_I, \Delta \log R_{23.5}) = (-1.5, 0.3)$ in the 3-D space.
This shift would result mainly from the adopted cosmology
($h=0.5$, $\Omega_0=1$), which could contribute to the shift in two ways:
(a) The $h$ shifts the observed galaxies through distance estimation.
If we change $h$ from 0.5 to 1, observed galaxies are shifted by
$(\Delta M_I, \Delta \log R_{23.5}) = (1.5, -0.3)$, which are sufficient
to explain the above shift.
(b) The lower $\Omega_0$ would increase the ratio of baryon to dark matter,
and then, decrease the mass-to-light ratio. If we decrease $\Omega_0$,
simulated galaxies would be shifted in the direction of $\Delta M_I < 0$
and $\Delta \log R_{23.5} > 0$. [Note on the contrary, if we assume a lower
baryon fraction in galaxies than the one adopted here, the simulated galaxies
would be shifted in the opposite direction.]

In our simulations, the procedure of galaxy formation and evolution
is not affected so much by changing the cosmology since we consider initial
conditions of nearly monolithic collapse. Hence the comparison
only in simulated galaxies would be possible even though the zero-point
is shifted.

\subsection{Origin of The Scaling Plane}
\begin{figure}
\plotone{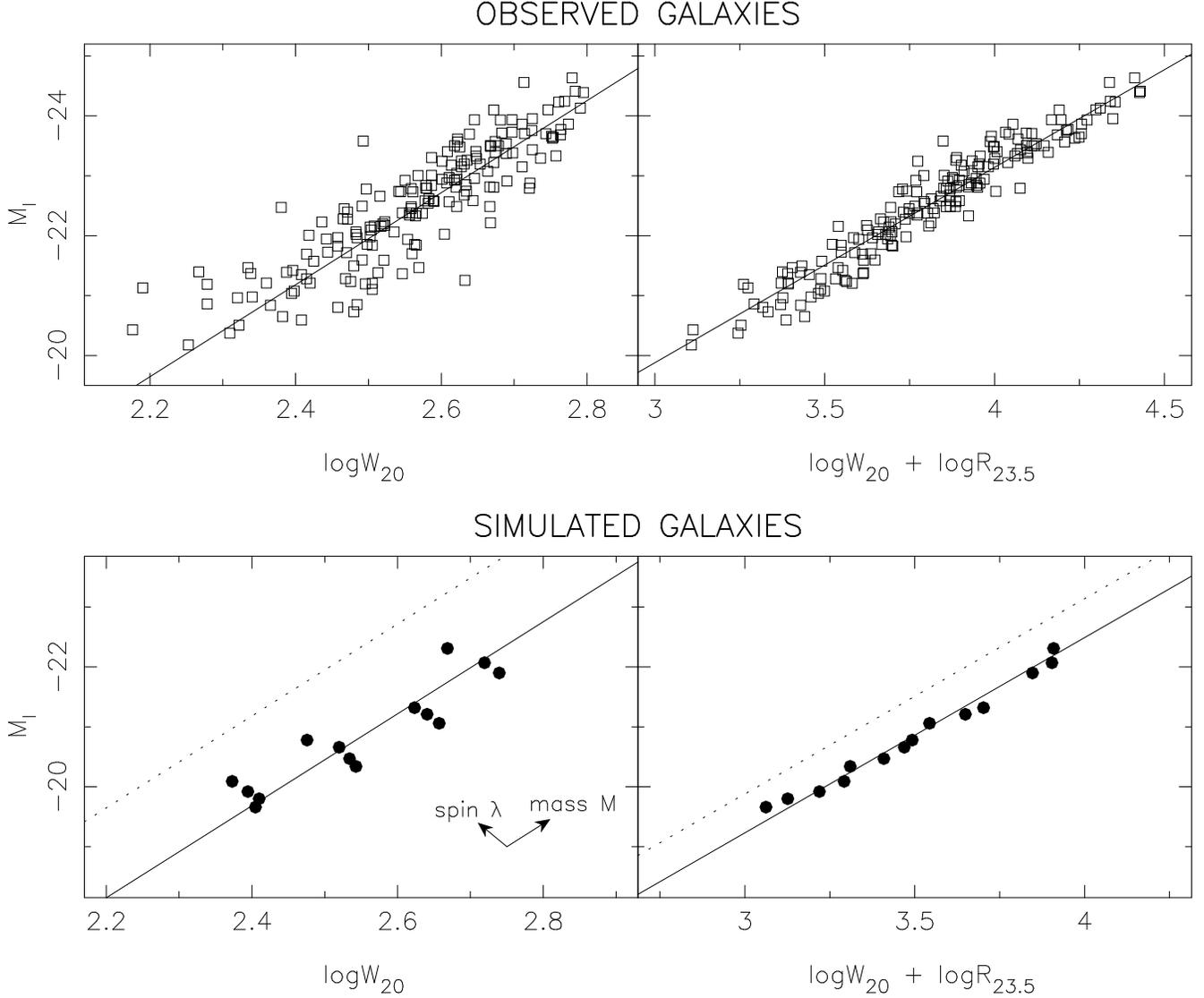}
\figcaption{
Comparison of observed (upper panels) and simulated (lower panels) galaxies in
the Tully-Fisher projection (left panels) and edge-on projection (right panels)
of the scaling plane.
The slopes of all the lines are determined by fitting to
the observation. In the lower panels, the dotted lines represent the observed
correlation (as do the solid lines in the upper panels), and the zero-points
of the solid lines are shifted by eye to fit them to the simulations.
The ranges of axes are different between upper and lower panels,
but the lengths of axes are exactly the same. Hence we can compare the slope
and scatter of the observations and the simulations.
In the Tully-Fisher projection of simulated galaxies (lower left),
the axes along mass and spin parameter are indicated by two arrows.
Comparing lower-left and lower-right panels, we find that the scaling plane
would originate from the difference of galactic mass and spin parameter.
\label{fig:comp}}
\end{figure}

As discussed in Section \ref{sec:obs}, the scaling plane has the primary
and secondary axes. Here we show that these two axes of the simulated
scaling plane correspond to galactic mass and angular momentum, respectively.
In order for these two parameters to correspond to the primary and secondary
axes exactly, they must satisfy the following three conditions:
(a) The axes along these two parameters are on the scaling plane.
(b) These axes are not parallel each other.
(c) The axis along mass (angular momentum) is parallel to
the primary (secondary) axis.
In Figure \ref{fig:comp}, the lower-right panel shows the edge-on
projection. All the simulated galaxies, which have different mass and
angular momentum, lie on the same scaling plane. The condition (a) is
satisfied. The axes along mass and angular momentum are illustrated
in the lower-left panel of Figure \ref{fig:comp}
(see also Koda, Sofue \& Wada 2000).
In this TF projection, the axes along mass and angular momentum
(spin parameter) are not parallel each other, which satisfies
the condition (b).
It is clear that the projections of the primary and secondary axes
onto the TF plot are along the directions of mass and angular momentum,
respectively, satisfying the condition (c).
We conclude that the scaling plane is spread by the difference of
primarily galactic mass and secondarily angular momentum.

The backbone of galactic scaling relations is the virial theorem.
Most of parameters would be determined on the domination of galactic mass.
However, if the mass is the only parameter which determines galactic
properties, galaxies would be distributed on a {\it line} in the 3-D space.
The secondary factor, spin parameter, causes a slight spread in properties
of disk galaxies. Then, spiral galaxies are distributed on a particularly
elongated (surfboard-shaped) plane in the 3-D space.

In fact, spin parameter (angular momentum)
affects galactic properties in the following three ways:
(i) Spin parameter changes the central concentration of disks in dark matter
halos. Lower spin parameter produces relatively concentrated disks and leads
to higher rotation velocities.
(ii) Spin parameter changes the radius of spiral galaxies.
Higher spin parameter produces galaxies with larger radii.
(iii) Therefore higher spin parameter produces galaxies with lower surface
densities, and then leads to slower star formation. It results in brighter
galaxies at $z=0$ because of the relatively younger age of their stellar
component. These three effects produce the scatters of the three scaling
relations ($L-V$, $V-R$, $R-L$).

\section{DISCUSSION}
We have introduced the scaling plane (unified scaling relations) of
observed spiral galaxies in the 3-D space of luminosity, radius and
rotation velocity, and investigated a possible origin of the scaling plane.
We have shown that mass primarily determines the galaxy position
in the 3-D space, and angular momentum (spin parameter) produces
a slight spread on the scaling plane.
The scaling plane is originated
in the galaxy formation process, controlled by these two factors,
mass and angular momentum.
In order to clarify the uniqueness of the origin, one could further consider
(1) other cosmological models (Mo, Mao \& White 1998),
(2) different ratios of baryon to dark matter,
(3) different mass aggregation histories (Avila-Reese et al. 1998),
and (4) other modelings of star formation and feedback (Silk 1997).

\begin{figure}
\plotone{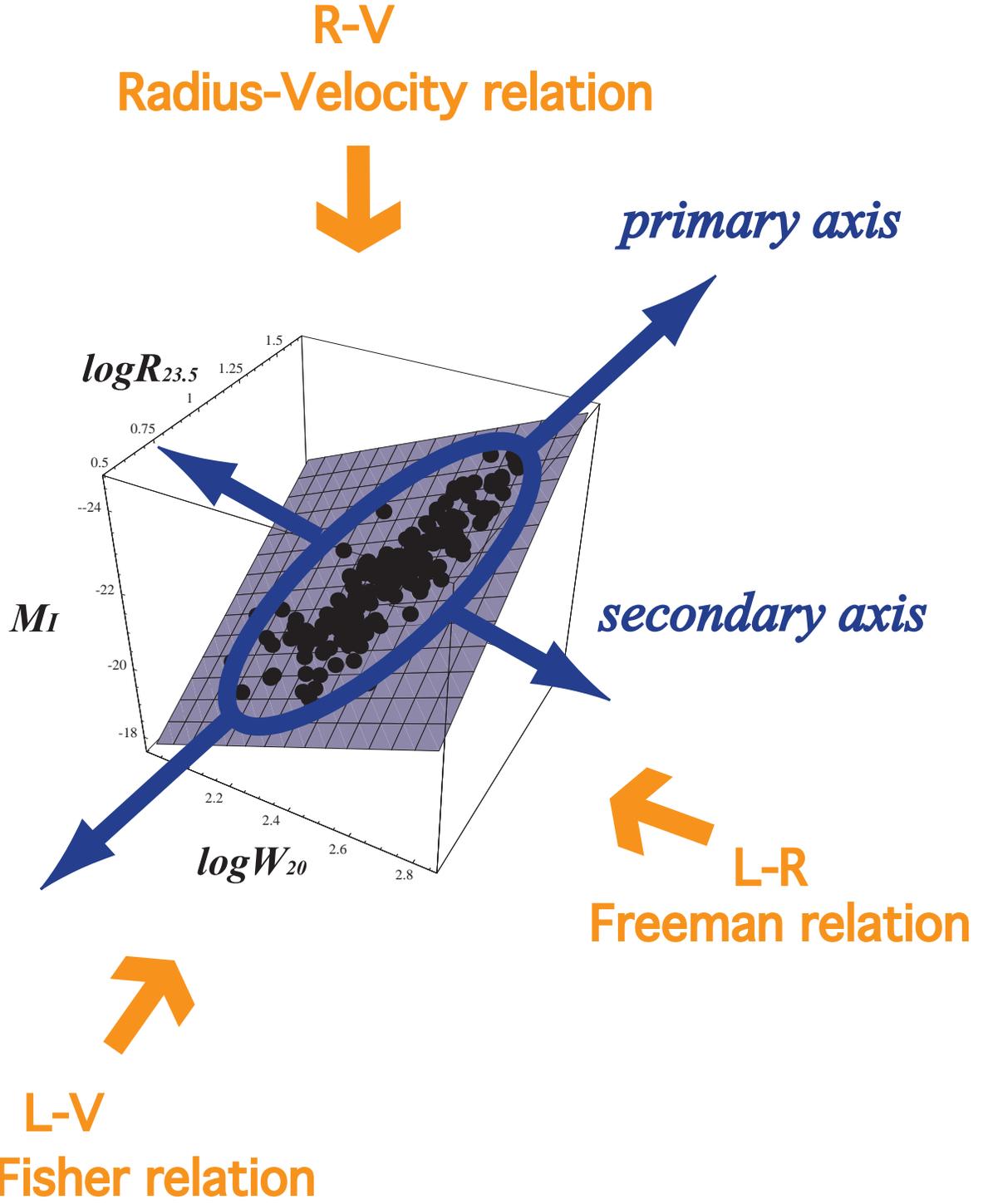}
\figcaption{
Observed spiral galaxies are distributed on a unique plane in the 3-D space
of luminosity $L$, radius $R$ and rotation velocity $V$,
hereafter {\it scaling plane}, and
distributed in a surfboard-shaped (particularly elongated) region
on the plane. In this schematic figure,
we use the $I$-band absolute magnitude $M_{\rm I} (\magni)$ for $L$,
face-on isophotal radius $R_{23.5} (\kpc)$ for $R$ and HI line-width
$W_{20} (\kmps)$ for $V$. The well-known scaling relations ($L-V$, $V-R$,
$R-L$) can be understood as oblique projections of the surfboard shape.
The scatters of these three correlations can also be unified by the scaling
plane. We hypothesize (i) that the 2-D distribution implies the existence of
two dominant physical factors in spiral galaxy formation,
and (ii) that one of them is more dominant than the other because of
the surfboard shape.
\label{fig:schem}}
\end{figure}

Many studies have concluded that there is no correlation of TF residuals
with radius and any other parameter. These results appear to be against
the existence of the scaling plane. We should note, however, that
the existence of the scaling plane does not imply a clear correlation
between TF residuals and radius, when the plane contains any kind of scatter,
e.g., observational errors or intrinsic one. The apparent discrepancy comes
from a confusion of two facts, that is, spiral galaxies are distributed
(i) on a plane, and (ii) in a surfboard-shaped region on it (see Section
\ref{sec:obs}). The definition of TF residuals are affected by the property
(ii). If the surfboard-shaped region rotates {\it on} the same plane,
the TF relation (proejcted relation) will be changed in its slope,
zero-point and 'the definision of residuals' as well
(cf. Figure \ref{fig:schem}). Hence the correlation of TF residuals with
radius is strongly affected by the property (ii), i.e., how galaxies are
distributed {\it on} the plane, and if the plane contains any kind of scatter
such as errors in observation, the combination of the property (ii) and the
scatter could hide the property (i), i.e., the existence of the scaling plane.

Still, the scaling plane implies correlations of each scaling relation
($L-V$, $V-R$, $R-L$) with surface brightness, at least in normal galaxies.
It is interesting to investigate whether low surface brightness (LSB)
galaxies are also distributed on the scaling plane. Zwaan et al. (1995)
discussed that LSB galaxies lie on the same TF relation as normal galaxies, 
while O'Neil, Bothun \& Schombert (1999) have concluded that their sample
of LSB galaxies does not produce the TF relation. So, the question is
still under debate, and further researches would be necessary to discuss
LSB galaxies in analyses of the scaling plane.
There have been studies which concluded that the Freeman's law would be
an artifact due to observational selection effects, because LSB galaxies
deviate from the luminosity-radius relation of normal galaxies (recently,
de Jong 1996; Scorza \& van den Bosch 1998). The scaling plane is so tight
that the plane itself would not be an artifact due to selection effects.
However, the galaxy distribution on the plane may change, if selection
effects affect the sample. LSB galaxies may provide a clue to understand
such selection effects, if they are the sequence of normal galaxies.

\acknowledgments
Numerical computations were carried out on the GRAPE system at the
Astronomical Data Analysis Center of the National Astronomical
Observatory, Japan. We would like to thank Dr. N. Arimoto for
providing us with the tables of the stellar population synthesis.
We are grateful to the anonymous referee for his/her fruitful comments.
J.K. thanks the Hayakawa Fund of the Astronomical Society
of Japan. J.K. also thanks Mrs. M. Redmond for reading the manuscript.

\clearpage

\end{document}